\documentclass{aa}
\usepackage{times,graphics}

\usepackage{epsfig} %
\begin{document}
   
   \title{Chemical composition of B-type supergiants in the OB8,
   OB10, OB48, OB78 associations of M31.}

   \subtitle{}

   \author{C. Trundle$^{1,2}$, P.L. Dufton$^{1}$, D.J.
   Lennon$^{2}$, S.J. Smartt$^{3}$, M.A. Urbaneja$^{4}$.}

   \offprints{C.Trundle at ct@ing.iac.es}
   
   \authorrunning{C.Trundle et al.}
   
   \titlerunning{Chemical composition of B-type supergiants in M31}
   
   \institute{The Department of Pure and Applied Physics,
              The Queen's University of Belfast,
	      Belfast BT7 1NN, Northern Ireland
		\and 
	      The Isaac Newton Group of Telescopes,
	      Apartado de Correos 321, E-38700,
	      Santa Cruz de La Palma, Canary Islands, Spain
	      	\and
	      Institute of Astronomy, University of Cambridge, 
	      Madingley Road, Cambridge, CB3 0HA
                \and
              Institutio de Astrof\'{i}sica de Canarias, E-38200
	      La Laguna, Tenerife, Spain
%             \thanks{}
             }

   \date{}

   \abstract{Absolute and differential chemical abundances are
   presented for the largest group of massive stars in M31
   studied to date. These results were derived from intermediate
   resolution spectra of seven B-type supergiants, lying within
   four OB associations covering a galactocentric distance of
   5$-$12\, kpc. The results are mainly based on an LTE analysis,
   and we additionally present a full non-LTE, unified model
   atmosphere analysis of one star (OB78-277) to demonstrate the
   reliability of the differential LTE technique. A comparison of
   the stellar oxygen abundance with that of previous nebular
   results shows that there is an offset of between $\sim 0.15 -
   0.4$\,dex between the two methods which is critically
   dependent on the empirical calibration adopted for the
   $R_{23}$ parameter with [O/H]. However within the typical
   errors of the stellar and nebular analyses (and given the
   strength of dependence of the nebular results on the
   calibration used) the oxygen abundances determined in each
   method are fairly consistent. We determine the radial oxygen
   abundance gradient from these stars, and do not detect any
   systematic gradient across this galactocentric range. We find
   that the inner regions of M31 are not, as previously thought,
   very 'metal rich'. Our abundances of C, N, O, Mg, Si, Al, S
   and Fe in the M31 supergiants are very similar to those of
   massive stars in the solar neighbourhood.} 

   \maketitle 

%
%________________________________________________________________

\section{Introduction} 
\label{intro} 

\begin{table*}
\caption[]{Observational details for the M31 targets.
The OB-association numbers are from van den Bergh (\cite{vnB64}), while the
stellar identifications are from Massey et al. (\cite{Mas86}). For example
OB8-17 is star number 17 in association 8. The alternative
identifications are from Berkhuijsen et al. (\cite{Ber88}). Visual
magnitudes and spectral types are taken from Massey et al. (\cite{Mas95}) --
the latter are identified by the authors initials, MAPPW. Also listed are  
the spectral types and heliocentric radial velocities, $v_r$ (in 
km s$^{-1}$), estimated from our spectra as discussed
in Sect. \ref{obsdata} and \ref{sptype}. The data for OB10-64 have
been previously discussed by Smartt et al. (\cite{Sma01b}).}
\begin{flushleft}
\centering
\begin{tabular}{llcllccc} \hline \hline

Star      &  Alternative  &   V    &   \multicolumn{2}{c}{Spectral Type}
& Time    & s/n & $v_r$ 
\\
Name      &      ID       &        &  MAPPW     & Here         	& (hours) & ratio &
\\
\hline   
\\
\\                                     
OB8-17    &  41-2178      & 18.01  &  O9-B1I    & B1Ia     	& 4.0	& 50 	& $-102\pm10$
\\    
OB8-76    &               & 18.52  &  B0III     & B0.5Ia	& 4.0  	& 30	& $-34\pm7$	 
\\
OB10-64   &  41-2265      & 18.10  &  B1I       & B0Ia          & 4.5   & 50    & $-113\pm11$
\\
OB48-234  &               & 18.50  &  B1I       & B1.5Ia 	& 2.5   & 30 	& $-125\pm8$	
\\ 
OB48-358  &               & 18.70  &  B0-1I     & B0Ia 	        & 2.5   & 30 	& $-107\pm22$
\\
OB78-159  &  40-1876      & 17.97  &  B0I	& B0Ia  	& 7.0   & 40	& $-467\pm20$
\\                              
OB78-277  &  40-1939      & 17.35  &  B1I	& B1.5Ia   	& 7.0   & 65	& $-567\pm9$
\\
OB78-478  &  40-2028      & 17.50  &  B0-1I     & B1.5Ia      	& 5.0   & 55	& $-561\pm12$
\\                              
\hline
\end{tabular}
\end{flushleft}
\label{obstable} 
\end{table*}

Until recently studies of the spatial distribution of chemical species
in M31 and many other external galaxies mainly involved H {\sc ii}
regions and supernovae remnants (SNR's) (see for example: Dennefeld \&
Kunth \cite{Den81}; Blair et al. \cite{Bla82}; Galarza et al.
\cite{Gal99}). In M31, the H {\sc ii} regions have low excitation,
thus the [O {\sc III}] lines used in estimating the electron
temperature are often too weak to be detected. Without a direct method
of determining the electron temperature, empirical calibrations must
be implemented and in some cases extrapolated to estimate abundances
of these H {\sc ii} regions. Depending on the empirical calibrations
adopted significantly different abundance estimates are derived (Pagel
et al. \cite{Pag80}; Mc Gaugh \cite{McG91}; Zaritsky et al.
\cite{Zar94}; Kobulnicky et al. \cite{Kob99}, Pilyugin \cite{Pil00},
\cite{Pil01a}, \cite{Pil01b}). This indicates that an independent
method of evaluating the abundance gradients of external galaxies is
needed. Blue supergiants are amongst the optically brightest stellar
objects in spiral and irregular galaxies, and can provide us with such a
method.

Previous studies have shown that quantitative spectroscopy of
B-type supergiants in Local Group galaxies can be carried out
successfully using 4 m telescopes (see Monteverde et al. \cite{Mon00},
Smartt et al. \cite{Sma01b}). The added advantage of using blue
supergiants  over H {\sc ii} regions is that their rich metal 
line optical spectra provide us with a means of studying elements not
observed in the emission line spectra of H\,{\sc ii} regions.
Other than H, He and CNO, some $\alpha$-processed and Fe-peak
elements are observed in the spectrum of blue supergiants. These
elements are important as they help to put constraints on 
nucleosynthesis theories and models of the chemical evolution of 
galaxies. 

Studies of blue supergiants in external galaxies can also help to
enhance the use of the Period-Luminosity (PL) relationship of
Cepheid variables in two ways. First one can determine {\em
stellar} abundances in the fields where Cepheids are found to
constrain the effect of metallicity on the PL relation. Secondly,
although it is difficult to determine extinction to Cepheids
themselves model atmosphere fits can accurately determine the
interstellar extinction towards blue supergiants, and hence
determine reddening in stellar fields where Cepheids are found. 
In addition, it now appears possible that blue supergiants could
be used to determine extragalactic distances by a properly
calibrated independent technique, using the wind
momentum-luminosity relationship (WLR, see Puls et al.
\cite{Pul96}; Kudritzki et al.  \cite{Kud99}; Kudritzki \& Puls
\cite{Kud00}). The WLR allows distances to be determined via
detailed studies of the radiatively driven winds in O, B and
A-type supergiants.  Several of the stars that will be presented
here have been observed by HST with both WFPC2 and STIS, in a
broader project to use them  as calibrators of the WLR within the
Local Group. First steps in this has occurred by  measuring the
terminal velocities from the UV STIS spectra in a related paper
(Bresolin et al. \cite{bres2002}). However to carry out the WLR
study in full the atmospheric parameters and chemical composition
of the early-type supergiants must be reliably estimated,  and
this is the focus of this paper. 

Previous to this work, the largest homogeneous abundance
analysis on a set of supergiants in M31 is that by Venn et al.
(\cite{Ven00} - hereafter VMLPKL). In this study, two A-type
supergiants and one F-type supergiant were analysed using high
resolution spectra from the Keck 10 m telescope and the HIRES
spectrograph. The stellar oxygen abundances in this analysis
suggested a shallow or negligible abundance gradient in M31. 
The nebular H\,{\sc ii} regions give  abundance gradients in the
range of -0.013 - -0.027 dex kpc$^{-1}$ , depending on the
empirical calibration used, however  the VMLPKL results were
based around only three stars and the individual abundances are
reasonably consistent to H\,{\sc ii} region results at similar
galactocentric distances. Definite conclusions from three points
are not warrented, and our experience should  caution us against
the use of restricted datasets in determining stellar abundance
gradients e.g. see the discussion of  Smartt \& Rolleston
(\cite{smr97}) on the abundance gradient in the  Milky Way.
Additionally Smartt et al. (\cite{Sma01b}) have analysed a B0Ia
supergiant, OB10-64, in the inner regions of M31 and compared it
to the abundances derived in the H\,{\sc ii} region surrounding
the OB10 association. The absolute nebular abundance is
critically dependent on which calibration of the $R_{23}$
parameter is used, however there is good agreement between the
stellar and at least one parameterization of the nebular result.

Here we present a spectroscopic analysis for the largest sample
of supergiants in M31 considered to date and covering a range in
galactocentric distance of 5 - 12 kpc. The elemental abundances
of these B-type supergiants are presented and their reliability
discussed including the uncertainties incurred by adopting an LTE
analysis. We also determine the oxygen abundance gradient in M31
and compare it with that determined from H\,{\sc ii} regions and
A \& F-type supergiants (VMLPKL).

%
%__________________________________________________________________

\section{Observational data}
\label{obsdata}

The double armed spectrometer ISIS on the William Herschel
Telescope was used to observe all the M31 stars during two
separate observing runs covering the nights 8-11 October 1997
and 28-29 September 1998 and exposure times are summarized in
Table \ref{obstable}. The targets designations are taken from
Massey et al.\ (\cite{Mas86}), with alternative identifications
from Berkhuijsen et al.\ (\cite{Ber88}) being given when
available. The spectrograph slit was positioned so that, if
possible it included two targets within a particular OB
association. 

The spectra presented here were obtained using only the blue
arm of the  spectrograph (which was fed with a beam folding
mirror) with the  R1200B grating and a EEV42-80 CCD (with
format 4096$\times$2048 13.5$\mu$m pixels). A slit
width of 1 arcsec was used giving a spectral resolution of
0.46 \AA\ /pixel. Although ISIS is a dual beam spectrometer
with the ability to gather simultaneous red and blue
spectra, the dichroic beam splitter introduces additional
structure into the blue spectra and reduces the throughput
of this arm. These ripples seriously compromise the quality
of stellar absorption line spectra, and hence we chose only
to observe in the blue. Spectra in the red region have been
taken at the Keck telescope and a full wind analysis of the 
H$\alpha$ region will be presented elsewhere.  The most
important lines for a {\em photospheric} analysis lie in the
blue which is the focus of  this paper. 

In Table \ref{obstable}, the total exposure times are listed and
as conditions varied significantly during the two observing
runs, the corresponding signal-to-noise ratio near to the
central wavelength of the blue spectra are also
given. Subsequently service spectroscopy was obtained for
Galactic supergiants covering a range of spectral types from
approximately B0Ia to B3Ia and these are summarized in
Table~\ref{obstand}. The instrumental setup was identical to
that used for the M31 targets with a mirror feeding the blue arm
of the spectrograph -- signal-to-noise ratios in the reduced
spectra were always in excess of 200 for these bright comparison
stars. 

The CCD frames were reduced to wavelength calibrated spectra
using the {\sc iraf} reduction system\footnote{{\sc iraf} is
written and  supported by the {\sc IRAF} programming group at
the National Optical Astronomy Observatories (NOAO) in Tucson
(http://iraf.noao.edu)}. Standard procedures were used to bias
correct and flat field  the stellar images. As our targets are
in  young clusters, there was often considerable  background
Balmer line emission from the nebular regions surrounding the
hot stars, which can vary over small spatial scales along the
slit. Our spectral resolution is easily high enough that the
important wings of H$\delta$ and H$\gamma$ will be unaffected
by the background nebulosity. However there may be residual
errors {\em in the cores} of the stellar Balmer line profiles
and less weight was given to modelling the line cores in the
analysis. The spectra were wavelength calibrated using Cu-Ne \&
Cu-Ar lamp exposures that interleaved the stellar spectra.
Individual stellar exposures were then cross-correlated to
search for significant wavelength shifts -- none were
identified. The spectra were then combined using {\sc scombine}
and a variety of rejection criteria such as rejecting pixels with
counts very high/low above/below the median; these were found to have
little effect on the final spectra.

The spectra were transferred to the {\sc starlink} spectrum 
analysis program {\sc dipso} (Howarth et al.\ \cite{How96}) 
for subsequent analysis. This included estimating the stellar
radial velocities and equivalent widths. Firstly metal line 
absorption features were identified in the spectra
of the M31 targets. For marginal features, the spectrum of 
a suitable Galactic comparison star (see Sect. \ref{sptype})
was used as a template and guide. The equivalent widths of the
features were then estimated in both stars by non-linear least
squares fitting of single or multiple Gaussian absorption
profiles to normalised spectra. Normally both the positions
and widths of the Gaussian profiles were allowed to vary. 
For the equivalent width estimates, the relative wavelength 
separations of lines within multiplets were fixed,
while for marginal features in the M31 targets, the widths were 
set to the mean value found for well observed lines. Further details
of these procedures can be found in Smartt et al.\
(\cite{Sma96}). In the case of OB10-64, the equivalent widths
were taken from Smartt et al. (\cite{Sma01b}).

For the radial velocity determinations, a common set of ten
relatively strong, isolated metal and helium lines were
adopted. For the M31 targets, the radial velocities,
transformed to a heliocentric frame are listed in Table
\ref{obstable}. The error estimates are the sample standard
deviation and if the errors are normally distributed the
uncertainty in the mean should be approximately a factor of
three smaller. The range in these error estimates probably
reflects the variations in the data quality and intrinsic line
strengths of the spectra. 

\begin{figure}
\resizebox{\hsize}{!}{\includegraphics{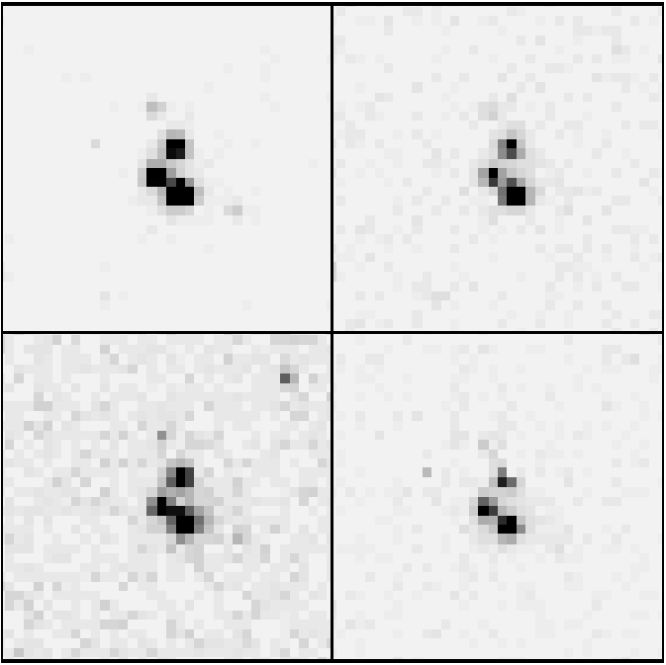}}
\caption[]{\label{ob78_159}WFPC2 images of OB78-159. These images
were taken in the filters U-B-I-V starting upper left and going 
clockwise. These images show that OB78-159 is a
part of a multiple system with at least 3 components. The size
of each image is 4*4 arcsecs.}
\end{figure}

Radial velocities of the OB48 stars and the OB78 stars, OB78-277
\& OB78-478, agree within 10\% of the radial velocities
determined by Rubin \& Ford (\cite{Rub70}) for H\,{\sc ii}
regions in the relevant OB associations. OB78-159 has a radial
velocity significantly less than that expected from a star
situated in the OB78 association. From WFPC2 images, taken in
the U, B, V and I filters, OB78-159 was observed to be part of a
multiple system (see Fig.~\ref{ob78_159}; from McCarthy et al.
in preparation),  which might  explain the difference in radial
velocity. An additional discrepancy is the large difference in
the radial velocities of the OB8 supergiants. No radial
velocities have been measured for H\,{\sc ii} regions in OB8,
but from a neutral hydrogen mapping of M31's velocity field the
radial velocity in the region of the OB8 association is expected
to be $-140 \pm 10$\,kms$^{-1}$ (Unwin \cite{Unw83}).
Furthermore, the OB10 association lies in close proximity to OB8
and the radial velocity of OB10-64 is $-$113 $\pm$11 kms$^{-1}$
(see Table~\ref{obstable}). Hence the radial velocity of OB8-17
agrees well with what we expect for a genuine single member of
the OB8  association. The radial velocity for OB8-76 differs by
some 70\,kms$^{-1}$ which could suggest the star is part of a
binary system, or has a peculiar velocities within the OB
association. Being part of a binary does not necessarily
exclude using the star in a detailed model atmosphere analysis
as the companion may contribute little or no flux at these
wavelengths, however it is an anomaly that should be kept in
mind during the analysis. Radial velocities were also estimated
for the comparison stars, the error estimates in this case being
smaller and typically $\pm$ 5 km s$^{-1}$. 

\begin{figure}
\resizebox{\hsize}{!}{\includegraphics{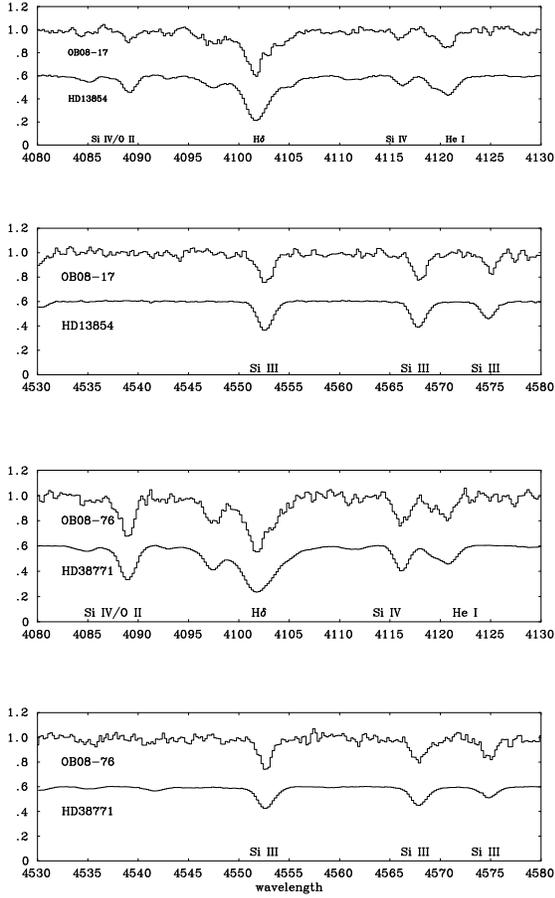}}
\caption[]{\label{m31_OB8}Spectra of selected wavelength regions
in two M31 targets, OB8-17 and OB8-76. The spectra show lines
of  \ion{Si}{iii} and \ion{Si}{iv} and the Balmer
H$\delta$\,line; the former  vary with effective temperature and
the latter with surface gravity.  Also shown are the spectra of
two Galactic supergiants, HD\,13854 (B1Iab)  and HD\,38771
(B0.5Ia), which appear to have similar spectral types. }
\end{figure}

%-----------------------------------------------------------------------
%
\section{Spectral types for the M31 targets}
\label{sptype}

From our spectra, we have estimated spectral types. Given the
relatively high spectral resolution and signal-to-noise, we expect
that these may be more reliable than those deduced previously.
Additionally this procedure will assist the identification of 
standards that will be suitable for a differential model 
atmosphere analysis.

For each of our M31 targets, our initial approach was to
attempt to identify a standard with both a similar hydrogen
line spectrum (which constrains the luminosity class) and
helium and metal line spectra (which constrains the spectral type).
The latter criterion implicitly assumes that both the M31 targets 
and the standards  have similar chemical composition. For some of our
M31 targets, the metal line spectra appeared to be weak (possibly
implying lower metal abundances) and here we have used the hydrogen
and helium spectra and the {\em relative} strength of the metal
lines. This is analogous to the procedures used by Lennon 
(\cite{Len97}) to estimate revised spectral types for SMC 
supergiants. The individual stars are discussed below:

\underline{OB8-17:} The observed spectra for this star has a relatively
high s/n ratio and, besides lines of neutral helium and hydrogen, has
a well developed metal line spectra. Comparison with the standard star
spectra showed good agreement for B1Ia and B1.5Ia, with the former being
probably superior. Selected regions of its observed spectra and 
that of the standard, HD\,13854, are shown in Fig. \ref{m31_OB8}.

\underline{OB8-76:} This star would appear to be hotter than OB8-17,
although the \ion{He}{ii} lines are not present in the observed spectrum.
Good agreement is found with the spectrum of HD\,38771 (B0.5Ia) and this is
again illustrated in  Fig. \ref{m31_OB8}. The spectrum of HD\,38771 may show
a very weak \ion{He}{ii} line at 4686 \AA\ but numerical simulations indicate
that this would not have been observable at the s/n ratio of the OB8-76
spectrum. Caution must be taken in adopting this spectral type
due to the possibility of OB8-76 being a composite star. 

\underline{OB10-64:} The spectral type of this star is B0Ia and is
discussed in Smartt et al. \cite{Sma01b}.

\begin{figure}
\resizebox{\hsize}{!}{\includegraphics{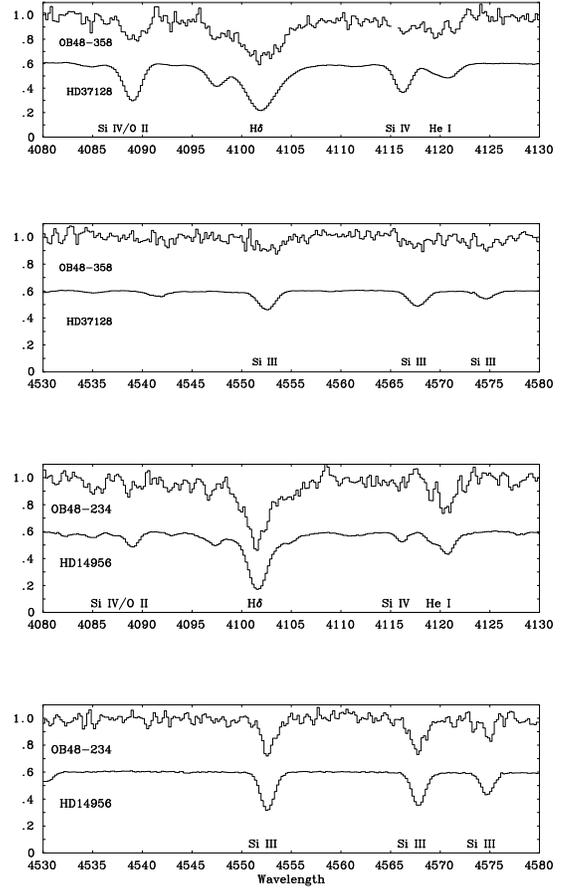}}
\caption[]{\label{m31_ob48}As for Fig.~\ref{m31_OB8}, however the spectra are
of the two M31 targets OB48-358 and OB48-234 and the two Galactic supergiants,
HD\,14965(B1.5Ia) and HD\,37128(B0Ia).  
}
\end{figure}

\begin{figure}
\resizebox{\hsize}{!}{\includegraphics{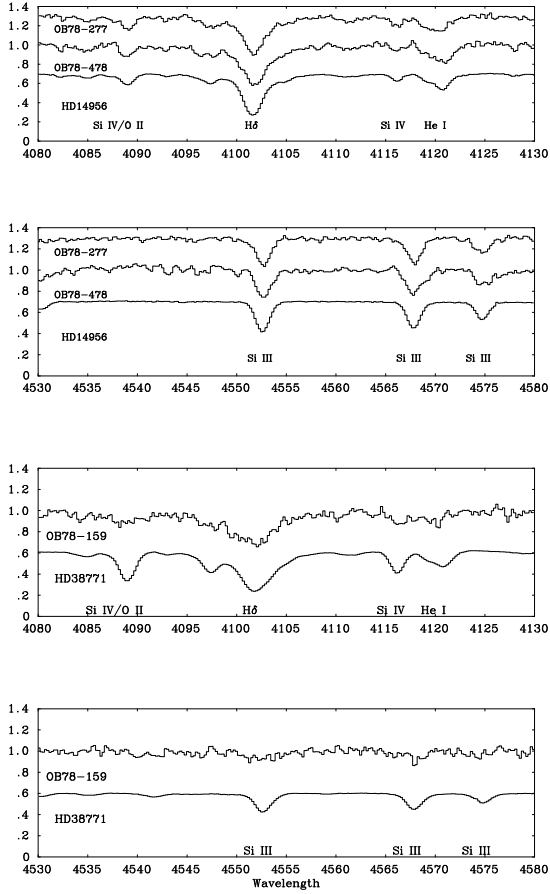}}
\caption[]{\label{m31_ob78}As for Fig.~\ref{m31_OB8}, however the spectra are of
three M31 targets -- two relatively cool (OB78-277 and OB78-478) 
and one relatively hot (OB78-159). 
The two Galactic supergiants are HD\,14956 (B1.5Ia) 
and HD\,38771 (B0Ia).
}
\end{figure}

\underline{OB48-358:} Again this star would appear to be hotter than 
the other target in the cluster. This leads to the \ion{He}{i} and
\ion{Si}{iii} spectra being relatively weak as can be seen in Fig.
\ref{m31_ob48}. From the hydrogen and diffuse helium lines, HD\,37128
would appear similar, although we emphasize that the spectral type
assignment may be subject to error.

\underline{OB48-234:} The spectra of both the OB48 targets appear to 
have weaker metal line spectra than those of their Galactic counterparts. 
In turn this, coupled with the relatively poor quality of the spectra, 
makes the assignment of spectral types more difficult. In Fig. 
\ref{m31_ob48}, the spectrum of OB48-234 is compared with that of the
standard HD\,14956 (B1.5Ia). There is reasonable agreement for the 
hydrogen and neutral helium lines but the silicon spectra is weaker
in the former - with the identification of the \ion{Si}{iv} lines
being marginal. Indeed a later spectral-type, such as B2Ia, would also
be compatible with the data.

\underline{OB78-277 and OB78-478:} These two M31 targets are considered
together as they have similar spectra, corresponding to spectral types 
of B1-2Ia. Indeed both spectra match well that of the Galactic standard
HD\,14956 (B1.5Ia) as can be seen from Fig. \ref{m31_ob78}. They
have therefore been assigned identical spectral types.
 
\underline{OB78-159:} This star appears to be hotter than the
other two  targets in the association. However surprisingly its
silicon spectrum appears to be significantly weaker than its
Galactic counterparts -- see Fig. \ref{m31_ob78}. This behaviour
is different to that found for both OB78-277 and OB78-478, which
are presumably relatively nearby objects. The weak metal line
spectrum makes the assignment of a spectral type more difficult
for OB78-159 but on the basis of the hydrogen and helium lines
(including the absence of identifiable \ion{He}{ii} lines) a
spectral type near B0Ia would appear appropriate. Moreover
OB78-159 appears to be part of a multiple system (see
Fig~\ref{ob78_159}) with three distinct point sources resolved by
HST, each of similar brightness in the $B-$band. Hence the
spectra is a composite of these three sources and is not a
single, B-type supergiant. It will not be considered further for
detailed analysis.

Two general comments can be made about the spectral-type
assignments. Firstly, there is relatively good agreement between
our spectral types  and those of Massey et al.\ (\cite{Mas95}),
although we are normally able to assign specific luminosity
classes and to narrow the range of possible spectral types to
one subgroup (i.e. B0, B1 or B2) -- this is not surprising given
the much higher spectral resolution and s/n ratios of our
spectra. Secondly, the stars would appear to fall into two
groups, five targets (OB8-17, OB8-76, OB10-64, OB78-277 and
OB78-478),
\begin{figure}
 \psfig{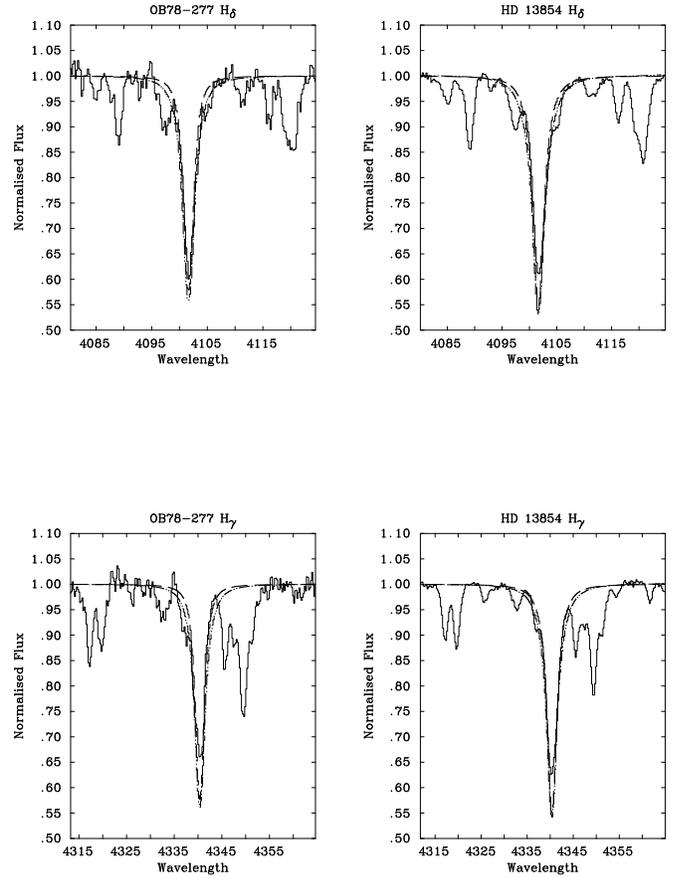}
\caption[]{\label{lteloggp}
The H$_{\delta}$ and H$_{\gamma}$ profiles for OB78-277 and
its Galactic counterpart, HD\,13854.  In the case of OB78-277
the theoretical profiles have surface gravities of $\log$ g =
2.5 \& 2.6, whereas for HD\,13854 the theoretical profiles are
for $\log$ g = 2.6 \& 2.7. The lower surface gravities are
represented by the dashed (- -) lines and the higher surface
gravities by the dot-dot-dashed (-$\cdot\cdot$-) lines. The
observed spectra are represented by the solid lines.
 }
\end{figure}
which have similar silicon line strengths to their
Galactic counterparts and two (OB48-234 and OB48-358), where the
silicon spectra are significantly weakened. It is plausible to
identify these differences as reflecting different metallicities
in the stellar atmospheres and in turn, in the interstellar
material from which they were formed. Indeed, considering the
galactocentric distance of these associations (see Table
~\ref{m31par}), the different metallicities would be consistent
with a decreasing radial gradient. We notice that OB78-159 has a
weaker silicon spectra than its standard and the other OB78
stars. 
However this does not imply that the OB78 association has
varying metallicity but is probably due to OB78-159 being a
composite (see Sect.~\ref{ltechem}).  We have also checked the 
HST WFPC2 images of the OB48 and OB78 associations which are
available and the stars OB48-234, OB48-358, OB78-277 and
OB48-478  all appear to be single point sources at this
resolution.  Photometry of these stars will be presented in a
future paper (McCarthy et al., in preparation). In addition both
OB8-17 and  OB10-64 have been observed by STIS (Bresolin et al.
\cite{bres2002}),  and the HST acquisition images again show
these are single point-like objects. Hence in the sample
considered here, we have only one object (OB8-76) which has not
been confirmed as a single object at HST's  spatial resolution
(at a distance of 783\,kpc, $0.1''$ corresponds to  0.4\,pc in
M31). 
\begin{table*}
\caption[]{Details of the Galactic comparison stars. Spectral types have been 
taken from Lennon et al.\ (\cite{Len92}), whilst the silicon and hydrogen line 
equivalent widths have been measured from our spectra. }
\begin{flushleft}
\centering
\begin{tabular}{llrrrrrrrr} \hline \hline
Star      &     Spectral & \multicolumn{2}{c}{\ion{Si}{ii}}&\multicolumn{3}{c}{\ion{Si}{iii}} &  \ion{Si}{iv}
\\
Name      &     Type	 & 4128\AA & 4130\AA & 4552\AA	& 4568\AA 	& 4574\AA 	& 4116\AA	
& H$\delta$	& H$\gamma$	 
\\
\hline   
\\           
HD\,167264  & O9.7Iab 	 & ...  &  ...        & 225 		& 200	        & 100	& 435       & 1680       & 1690
\\
HD\,37128   & B0Ia	 & ...  &  ...        & 304		& 243 		& 113	& 489       & 1630	 & 1560
\\
HD\,38771   & B0.5Ia  	 & ...  &  ...        & 376		& 314		& 182	& 357       & 1810	 & 1810
\\
HD\,2905    & BC0.7Ia	 & ...  &  ...        & 395		& 350		& 219	& 228       & 1360	 & 1440
\\
HD\,13854   & B1ab	 & ...  &  ...        & 456		& 390		& 241	& 135       & 1600       & 1640
\\
HD\,14956   & B1.5Ia	 & ...  &  ...        & 494		& 426		& 280	& 94        & 1390	 & 1470
\\
HD\,190603  & B1.5Ia+    & 42&  72      & 485 		& 438 		& 293   &114        & 1130	 & 1170
\\
HD\,14143   & B2Ia	 & ...  &  ...        & 463		& 394		& 248	& 51        & 1320	 & 1280
\\
HD\,14818   & B2Ia	 & ...  &  ...        & 428		& 368		& 224	& 49        & 1550	 & 1570
\\
HD\,194279  & B2Ia	 & ...  &   ...       & 468		& 415		& 272	& 69        & 1130	 & 1260
\\                              
\hline
\end{tabular}
\end{flushleft}
\label{obstand} 
\end{table*}

%-----------------------------------------------------------------------
%
\section{LTE model atmosphere and abundance analyses}
\label{ltemod}

All the M31 targets are luminous supergiants and as, for
example, discussed by McErlean et al. (\cite{McE99} - hereafter
MLD) their spectra will be affected by significant non-LTE
effects. However MLD also showed that even with a non-LTE
approach based on plane-parallel models without line-blanketing 
and stellar winds, it was difficult to obtain reliable {\em
absolute} atmospheric parameters and abundances. Hence we will
initially adopt a simple approach and use differential LTE
methods. This will inevitably lead to the estimates of the
atmospheric parameters and absolute chemical composition having
significant uncertainties. However provided the M31 targets are
well matched (in terms of their atmospheric parameters) to the
Galactic standards, differences in atmospheric parameters and
chemical compositions should be more reliably estimated. In Sect
5. we will carry out non-LTE calculations and discuss the
effects departure from LTE conditions has on the differential
abundances of these supergiants.

%-----------------------------------------------------------------------
%
\subsection{Atmospheric parameters} 
\label{lteatpar}

\begin{table*}
\caption[]{Silicon and hydrogen line equivalent widths for the M31 targets}
\begin{flushleft}
\centering
\begin{tabular}{llrrrrrrrr} \hline \hline

Star      &\multicolumn{2}{c}{\ion{Si}{ii}} &\multicolumn{3}{c}{\ion{Si}{iii}} &  \ion{Si}{iv}
\\
 Name     & 4128\AA & 4131 \AA &4552\AA	& 4568\AA 	& 4574\AA 	& 4116\AA	
& H$\delta$	& H$\gamma$	 
\\
\hline                 
\\
OB8-17   & ...   & ...    &476      	& 392		& 241		& 116  	& 1350		& 1370
\\  
OB8-76   &  ...  &  ...   &391	& 341		& 256		& 414   & 1780		& 1600
\\
OB10-64  &  ...  &  ...   &245        &240            &120            & 400   &  1690       & 1580 
\\
OB48-234  & 97 & 126 &457	& 426		& 240		&   ...    & 1620		& 1500
\\
OB48-358  & ...   &  ...   &404	& 284		& 300:		& 221:  & 1600		& 1580
\\
OB78-159  & ...   &   ...  &293	& 237		& 119:		& 141   & 1640		& 1540
\\
OB78-277  & ...   & ...    &473 	& 438		& 269		& 164   & 1470		& 1380
\\
OB78-478  & ...   &  ...   &487	& 466		& 316		& 101   & 1650		& 1750
\\                              
\hline
\end{tabular}
\end{flushleft}
\label{m31sih} 
\end{table*}

The LTE methods employed are similar to those described in Smartt et
al. (\cite{Sma96b}) and Rolleston et al. (\cite{Rol97}),  where more
details can be found. All results are based on the grid of unblanketed
LTE model atmospheres generated using the code TLUSTY (Hubeny
\cite{Hub88}). This grid covered a range in effective temperatures of
10,000 $\leq$ T$_{\rm eff}$ $\leq$ 35,000 and included logarithmic
surface gravities from $\log$ g = 4.5 down to near the Eddington
stability limit. LTE line formation codes were then used to derive
line profiles and equivalent widths leading to estimates of
atmospheric parameters and chemical compositions. 

Although this approach appears to be attractive, for evolved
stars it has one major disadvantage. This is that it implicitly
assumes that the chemical composition of the standard is
representative of that of the Galactic ISM. However for
supergiants, surface chemical compositions may be affected by
material mixed from the stellar core that has been processed by
nuclear reactions. Indeed MLD found evidence that the nitrogen
and carbon abundances could be significantly modified by such
mixing. Hence we will use non-LTE techniques in Sect.~\ref{nlte} to
attempt to estimate more reliable absolute parameters for our M31
targets.

\begin{table*}[t]
\caption[]{LTE atmospheric parameters and galactocentric
distances for both the M31 and
Galactic standard stars. The random errors in the atmospheric
parameters are  $\Delta$T$_{\rm eff}$ = $\pm$ 1,000 K;
$\Delta$$\log$ g = $\pm$ 0.2 dex; $\Delta\xi = \pm$ 5 kms$^{-1}$;
$\Delta$v $\sin$ i = 15 kms$^{-1}$. Also included is the degree of
CN processing in the stars, the classifications are taken from MLD
as are the designations for the Galactic standards. The status of
the M31 stars are discussed in Sect.~\ref{chempro}. ($\times$ -
'normal/moderate'; $\circ$ - 'processed?'; $\bullet$ -
'highly-processed'). The galactocentric distances of the Galactic
standards are taken from Humphreys (\cite{Hum70})}.  
\begin{flushleft}
\centering
\begin{tabular}{lrcccrc} \hline \hline
Star Name      & T$_{\rm eff}$ (K)&  $\log$ g (dex) & $\xi$ & v $\sin i$
(kms$^{-1}$) & R$_{g}$ (Kpc) & Processing\\
\hline   
\multicolumn{6}{c}{M31 Stars}
\\ 
 \hline   
\\            
OB8-17   & 22,000  & 2.5   & 32  & 72  &  5.85 & $\bullet$	    
\\
OB8-76   & 27,500  & 3.0   & 19  & 79  &  6.12 & $\circ$    
\\
OB10-64   & 30,500  & 3.4   & 21  & ...  & 5.90 & $\circ$
\\
OB48-234  & *21,500 & 2.5   & 27  & 64  & 11.41 &$\times$	    
\\
OB48-358  & 26,000  & 2.9   & 21  & 70  & 11.59	& $\times$    
\\
OB78-159  & 26,000  & 3.0   &  ... &  ... &  9.50	    
\\
OB78-277  & 23,000  & 2.5   & 24  & 73  &  9.58	&$\circ$    
\\
OB78-478  & 20,500  & 2.4   & 24  & 86  &  9.06	&$\bullet$    
\\                              
\hline
\multicolumn{7}{c}{Galactic Standards}
\\ 
\hline          
\\ 
HD\,167264  & 31,500  & 3.5   & 20  & ...  & 8.73 &$\circ$
\\         
HD\,38771   & 29,500  & 3.2   & 25  & 90  & 10.37 &$\circ$	      
\\
HD\,2905    & 25,500  & 2.8   & 21  & 79  & 10.48 &$\times$	      
\\
HD\,13854   & 23,000  & 2.6   & 27  & 64  & 11.30 &$\bullet$	      
\\
HD\,190603  & 21,000  & 2.3   & 28  & 70  &  9.69 &$\bullet$	      
\\
           & *21,500 & 2.3   & 23  & 70  & & 	      
\\
HD\,14956   & 20,500  & 2.3   & 30  & 85  & 11.33 &$\bullet$	      
\\
HD\,14818   & 18,000  & 2.2   & 35  & 63  & 11.59 &$\bullet$	      
\\ 
\hline          
\multicolumn{7}{l}{\footnotesize{(*) The effective temperature
for this star is determined from the Si {\sc ii}/Si {\sc iii} ionisation}}\\
\multicolumn{7}{l}{\footnotesize{stages
rather than the Si {\sc iii}/Si {\sc iv}.}}     
\end{tabular}
\end{flushleft}
\label{m31par} 
\end{table*}

The atmospheric parameters were initially estimated from the Si
{\sc iii}/Si {\sc iv} ionisation balance and the fitting of
hydrogen line profiles to the observed spectra. The former
principally constrains the effective temperature and the latter,
the surface gravity. It is important to note that these effective
temperatures are derived from unblanketed model atmospheres which
may affect the ionisation balance and hence the absolute
abundances. In order to derive highly accurate {\it absolute}
abundances line-blanketed model should be used, this again
stresses the importance of the differential analysis as any
systematic errors arising from using unblanketed models will not
affect the differential analysis. In the case of OB48-234 the
Si~{\sc ii}/Si {\sc iii} ionisation balance was utilised due to
the absence of the Si {\sc iv} line in the observed spectrum. 

The quality of the fits for the hydrogen line profiles is illustrated
in Fig.~\ref{lteloggp}. The surface gravity is based on the fitting of
the profile wings as the line cores are contaminated by both non-LTE
effects and by wind emission (see Sect.~\ref{nlte}). Both the H$_{\delta}$ and H$_{\gamma}$
features are significantly blended with metal lines. Hence more weight
was given to the long wavelength part of the profile in the former
case and to the short wavelength part of the profile in the latter
case. The equivalent widths of the silicon and hydrogen lines in the
Galactic standards and M31 supergiants are shown in
Tables~\ref{obstand} \& \ref{m31sih}.

Microturbulence, $\xi$, also needs to be included in the
iterative process to determine the atmospheric parameters. This
parameter is introduced to ensure no variation of abundance with
line strength within a particular ionic species.  
The microturbulent velocity of each star in the
sample was determined from both the O {\sc ii} and Si {\sc iii}
lines. Although O {\sc ii} has the richest line spectrum in these
early B-type stars, with $\sim$ 20 unblended lines in the program
stars and $\sim$ 35 in the Galactic standards, the scatter in
abundances from one multiplet to another, due to errors in the
atomic data and possible non-LTE effects, makes it difficult to
constrain the appropriate value of microturbulence. On the other
hand, the Si {\sc iii} multiplet at 4560 \AA\ has three lines
covering a significant range in line strength and being from the
same multiplet the abundances derived from these silicon lines
should be less affected by such uncertainties than the oxygen
lines. Hence the microturbulent velocities listed in
Table~\ref{m31par} are derived from the silicon lines. It is
interesting to note that in all cases the microturbulent
velocities determined from the O {\sc ii} lines were in the range
of 5 - 7 kms$^{-1}$ larger than those derived from the Si {\sc
iii} multiplet. This effect has also been observed in non-LTE
analyses of early B-type giants and supergiants by Vrancken et
al. (\cite{Vra00}) and MLD, respectively. 

As can be seen from Table \ref{m31par} the microturbulence determined
for both the M31 and Galactic supergiants are in the range of 19 to 35
kms$^{-1}$, indicating supersonic velocities. For the Galactic
standards, they are on average 15 kms$^{-1}$ greater than the values
determined by MLD in their non-LTE analysis of the same sample. This
discrepancy in microturbulence from LTE and non-LTE analyses has also
been observed by Gies \& Lambert (\cite{Gie92}), in a sample of 5
supergiants. Since adopting a non-LTE analysis reduces the value of
$\xi$, the large value we observe is probably due to departures from
LTE (see Smartt et al. \cite{Sma97}). It may also be due to the presence of a
macroscopic velocity field which can mimic microturbulence (see
Kudritzki et al. \cite{Kud92}). However, as the effect is seen in both program
and standard stars the differential abundances should still be
reliable.

In the case of OB48-358, only an upper limit in the equivalent
width of Si {\sc iii} 4574\AA\ was obtained thus the
microturbulence could not be derived from the Si {\sc iii}
multiplet. However, we have adopted the same microturbulence as
that of its standard HD\, 2905, $\xi$=21 kms$^{-1}$, this is in close
agreement to that derived from the O {\sc ii} lines of OB48-358,
which is 20 $\pm$ 7 kms$^{-1}$.

The projected rotational velocity, $v \sin i$, can be estimated
when analysing supergiant spectra. This line broadening parameter
was determined from a selection of seven unblended metal lines by
fitting theoretical profiles to the observed spectra. The
theoretical profiles were convolved with a gaussian profile to
account for the instrumental broadening. The profile is then
convolved with a rotational line profile fourier transform
function, this has been described in detail by Gray (\cite{Gra76}) and
more recently by Rucinski (\cite{Ruc90}). It should be noted that the
spectra of early B-type supergiants appear to show significant
broadening (see, for example, Howarth et al. \cite{How97}). Hence, these
estimates probably reflect the effects of both rotation and
turbulence. The rotational velocity estimates are listed in Table
4 along with the adopted values for the above mentioned
atmospheric parameters.

\subsection{Chemical composition}
\label{ltechem} 
\begin{table*}
\caption[]{LTE absolute elemental abundances for the M31
B-type supergiants given as $\log[\frac{X}{H}]$ + 12 (where X=element). The
errors represent the standard deviation of the mean. }
\begin{flushleft}
\centering
\begin{tabular}{rlllllll} \hline \hline
Ion         & OB8-17           &  OB8-76           &         OB10-64   &         OB48-234  &         OB48-358  &         OB78-277  &  OB78-478  \\ \hline
He {\sc i}  & 10.28 $\pm$ 0.28 & 11.50 $\pm$ 0.17 & 11.44 $\pm$ 0.22 & 10.97 $\pm$ 0.23 & 11.12 $\pm$ 0.33 & 10.89 $\pm$ 0.32 & 11.08 $\pm$ 0.37 \\   
C {\sc ii}  & 7.45             & $<$7.80          & 8.09             &  7.68            &  $<$7.36         &  7.50            & $<$7.69          \\       
N {\sc ii}  & 8.24 $\pm$ 0.19 &  8.92 $\pm$ 0.35  & 8.13             &  8.19 $\pm$ 0.16 &  $<$7.69         &  8.09 $\pm$ 0.13 & 8.51 $\pm$ 0.18  \\     
N{\sc iii}  & ...             &  8.31             & 8.36 $\pm$ 0.21  & ...              &   ...            &  ...             & ...   \\
O {\sc ii}  & 8.44 $\pm$ 0.17 &  9.07 $\pm$ 0.26  & 8.74 $\pm$ 0.24  &  8.52 $\pm$ 0.22 &  8.66 $\pm$ 0.44 &  8.78 $\pm$ 0.38 & 9.02 $\pm$ 0.24  \\   
Mg {\sc ii} & 7.60            &  $<$8.39          &  7.62            &  7.61            &  $<$7.37         &  7.53            &  7.42            \\
Al {\sc iii} & 7.35            &  $<$7.76         &  ...             & 7.24             &  $<$7.66         &   6.97           & 7.68            \\  
Si {\sc ii}  & ...             &  ...             &  ...             & 7.50 $\pm$ 0.03  &  ...             &  ...             & ...             \\  
Si {\sc iii} & 7.26 $\pm$ 0.01 &  8.17 $\pm$ 0.03 &  8.04 $\pm$ 0.10 & 7.47 $\pm$ 0.10  &  7.42 $\pm$ 0.17 &  7.49 $\pm$ 0.06 & 7.90 $\pm$ 0.07 \\     
Si {\sc iv}  & 7.34            &  8.29            &  7.96           & 7.33             &  7.47                   &  7.43 & 7.89\\     
S {\sc iii}  & 6.99 $\pm$ 0.08 &  7.53            &  ...             & 7.08             &   $<$6.92        &  6.91            & 7.36 $\pm$ 0.23 \\      
Fe {\sc iii} & 7.28            &  ...		  &  ...             & $<$6.98          &  ...             &  ...             & 7.22            \\ 
\hline                
\end{tabular}
\end{flushleft}
\label{m31ab} 
\end{table*}

\begin{table*}
\caption[]{LTE line-by line differential elemental
abundances for the M31 supergiants given as $\Delta
\log[\frac{X}{H}] = \log[\frac{X}{H}]_{*} - \log[\frac{X}{H}]_{standard}$.
The Galactic standards that were used in the differential
analysis are listed in brackets under the appropriate M31
star.}
\begin{flushleft}
\centering
\begin{tabular}{rlllllll} \hline \hline
             & OB8-17           &  OB8-76           &         OB10-64   &         OB48-234  &         OB48-358  &   OB78-277  &  OB78-478  \\
 Ion         & (HD\,13854)      & (HD\,38771)      & (HD\,167264)     &  (HD\,190603)       & (HD\,2905)        & (HD\,13854)  & (HD\,14956)       \\  \hline
He {\sc i}   & $-$0.12 $\pm$0.28  & +0.34 $\pm$0.46 &  +0.04 $\pm$0.20 & $-$0.06 $\pm$0.14 & +0.05 $\pm$0.23  &  $-$0.18 $\pm$0.30 & $-$0.05 $\pm$ 0.13 \\
C {\sc ii}   & +0.20            & $<$0.20         &  $-$0.14           & +0.41           & $<-$0.08         &  +0.21           & $<$0.33          \\
N {\sc ii}   & +0.16 $\pm$0.12  & +0.72 $\pm$0.36 &  $-$0.09           & $-$0.24 $\pm$0.30 & $<$0.09          &  +0.02 $\pm$0.09 & +0.05 $\pm$ 0.21  \\  
N{\sc iii}   & ...              & +0.30           &  $-$0.04 $\pm$0.29 & ...             & ...              &   ...            &    ...           \\
O {\sc ii}  & $-$0.05 $\pm$0.14  & +0.42 $\pm$0.25 &  $-$0.02 $\pm$0.22 & $-$0.10 $\pm$0.43 & $-$0.08$\pm$0.47   & +0.21 $\pm$0.19  & +0.51 $\pm$ 0.37  \\
Mg {\sc ii}  & +0.15            & $<$0.40         &  $-$0.01           & $-$0.02           & $<-$0.17         & +0.08            & +0.03             \\
Al {\sc iii} & +0.16             & $<$0.55         & ...              & +0.07           & $<$0.45          &  $-$0.24           & +0.39             \\
Si {\sc ii}  &  ...             & ...             &  ...             & $-$0.08 $\pm$0.09 &  ...             &      ...         & ...                \\
Si {\sc iii} & $-$0.04 $\pm$0.02  & +0.45 $\pm$0.02 &  $-$0.15 $\pm$0.09 & $-$0.16 $\pm$0.07 &  $-$0.23 $\pm$0.26 & +0.19 $\pm$0.05  &  +0.36 $\pm$ 0.07 \\
Si {\sc iv}  & -0.01 & 0.62 &  -0.21  & -0.25           & +0.03  & +0.08  & +0.40\\
S {\sc iii}  & $-$0.06 $\pm$0.02  & +0.13           &  ...             & +0.05           &  $<-$0.11        & $-$0.15            & +0.12  $\pm$ 0.15  \\ 
Fe {\sc iii} & +0.41            &  ...            &  ...             & $<-$0.07        &  ...             &     ...          & +0.32              \\
\hline                
\end{tabular}
\end{flushleft}
\label{m31diff} 
\end{table*}

In this section we will present the results of the LTE abundance
analysis for seven of the M31 B-type supergiants and discuss the
abundances of individual elements. Although it was possible to
make an approximate estimate of the effective temperature and
surface gravity of OB78-159, its composite nature excludes it
from a quantitative abundance analysis and it is therefore
omitted from any further discussion. The quality of the OB8-76
spectrum was such that a differential abundance analysis was
possible, and the results appear to be consistent and reasonable.
However they should be treated with caution as in the light of
its peculiar radial velocity this target may be a binary.

As mentioned at the beginning of Sect.~\ref{ltemod}, we will
concentrate on the results of differential abundance analysis rather
than the absolute analysis, the results of which can be seen in
Table~\ref{m31ab} \& \ref{m31diff}. For the differential analysis we
attempted to fulfill certain criteria in selecting the appropriate
standards for each target. The criteria were: 1) To match, as closely
as possible, the atmospheric parameters and spectral type, 2) To only
use those stars that had unprocessed atmospheres according to MLD.
Although it was possible to fulfill the first of our criteria for all
cases, the second  criterion was more difficult to adhere to, with it
only being possible for OB48-358. In the cases of the  four other
targets only Galactic supergiants which were described by MLD as
`highly processed', have similar atmospheric parameters. According to
MLD, the identifier `highly processed' represents a supergiant with a
photosphere contaminated by the products of CN-cycle burning and so
should exhibit a significant nitrogen enhancement with a lesser
carbon depletion. In Sect.~\ref{chempro}, we will discuss the effect
that the possible contamination of the photospheres of our standards
has on our analysis.   

{\bf Helium:} The helium abundance was determined from the
non-diffuse He {\sc i} lines at 3965, 4121
and 4713\AA\ which were well observed features in most of our M31
targets. However for the cooler stars  the 4121\AA\
line tended to be blended with metal lines making the line
strengths difficult to determine. Weak features were observed in
OB8-17 and OB48-234 at 4437\AA , while a weak line was also observed
in OB8-17 at 4169\AA . In addition to the non-diffuse lines,
helium abundances were determined from the diffuse lines, this
was done using profile fitting techniques. In all cases the
results from the diffuse and non-diffuse helium lines were
consistent. From Table ~\ref{m31diff} we see that the helium
abundance of each of these stars is consistent with that of the
standard stars, {\it within the errors}.

{\bf Carbon \& Nitrogen:} The carbon abundances are determined
from the close doublet at 4267\AA\ . This feature is known to be 
difficult to model in both LTE and non-LTE and hence the
absolute abundances from this line should be treated with caution
(see Lennon \cite{Len83}; Eber \& Butler \cite{Ebe88}; Sigut
\cite{Sig96}; MLD). The CII 4267\AA\
feature in OB8-17 appeared to be affected by noise which may
account for it having a larger abundance than its Galactic
counterpart. Unlike carbon, nitrogen has a relatively rich
spectrum in the M31 stars except for the hotter stars OB8-76
and OB48-358 where the lines 3995\AA, 4236\AA\ and 4241\AA\
are the best observed lines. These two elements are
important for identifying which stars have nucleosynthetically
processed material mixed to their stellar surface.

{\bf Oxygen:} This element has the richest spectrum in these
early B-type supergiants and is a powerful diagnostic tool
for determining the abundance gradient of M31 (see Sect.~\ref{discuss}).

{\bf Silicon:} As discussed previously in Sect.~\ref{lteatpar}, Si {\sc iii}
\& Si {\sc iv} were used to determine the atmospheric parameters.
The Si {\sc iv} abundances presented in Table~\ref{m31ab} are
based on the 4116\AA\ line except for in the case of OB48-234,
where this line was unobserved in the spectrum. The Si {\sc
iv} abundance of OB48-234 is based on the 4088\AA\ line. The absolute
abundances derived from the 4088\AA\ line in the other six stars tended to be larger
than that from the 4116\AA\ line. Although blending with a nearby
O {\sc ii} line was taken into account when measuring
the equivalent width of this silicon line there is evidently
still some contribution from this oxygen feature.

{\bf Magnesium, Sulphur, Aluminium:} The magnesium abundances are
based on a single feature, this is the close doublet at 4481\AA\
. This feature was well observed in most of the targets except
OB8-76 \& OB48-358, where it was only possible to determine upper
limits. From the differential analysis the magnesium abundance in
the target stars appear to be similar to that of the standards.
The close doublet of Al {\sc iii} at 4529\AA\ is the only
unblended aluminium feature in these early B-type supergiants.  Again only upper
limits could be derived for both OB8-76 \& OB48-358. The sulphur
abundance is determined from the S {\sc iii} feature at 4285\AA
, where only an upper limit could be determined for OB48-358. In
the cooler stars of the sample OB8-17 and OB78-478 an additional
S {\sc iii} feature was observed at 4361\AA .

{\bf Iron:} The iron abundance was derived from a single Fe
{\sc iii} feature at 4419\AA . This feature was only observed
in the cooler stars of the sample OB8-17, OB48-234 \&
OB78-478. From Table~\ref{m31diff} it seems that iron is
overabundant in OB8-17 \& OB78-478. However this feature is
relatively weak and in the standard spectra is affected by a
diffuse interstellar absorption band. Hence this enhancement may
not be significant.

%-----------------------------------------------------------------------
%
\section{Non-LTE model atmosphere and abundance analyses}
\label{nlte}

A criticism to taking the LTE approach when analysing
supergiants is that the derived abundances may be subjected to
significant non-LTE effects. In order to investigate the
uncertainties incurred in adopting a LTE analysis we have also
carried out a non-LTE analysis on one of the M31 supergiants and
its Galactic standard. Given the relatively high signal-to-noise of
OB78-277, it was decided to carry out the non-LTE analysis for
this star and its Galactic counterpart, HD\,~13854.

The non-LTE code used is that by Santolaya-Rey, Puls \& Herrero
(\cite{San97}). This is a "unified model atmosphere" code, which
includes spherical extension and stellar winds, although omits metal
line-blanketing at the present stage (see Santolaya-Rey et al.
\cite{San97} for full details of code). This code calculates both the
atmospheric structure and line profiles and has a similar philosophy
to that of the line formation code DETAIL (Butler \& Giddings
\cite{But85}), i.e the code is data driven. The 
code includes atomic data for hydrogen, helium, carbon,
nitrogen, oxygen,magnesium  and silicon. The atomic models and Stark
broadening data implemented in the line formation of hydrogen and
helium are described in full in Santolaya-Rey et al. (\cite{San97}). In the
case of silicon, nitrogen and oxygen the metal ion populations and
line profiles were calculated using the atomic data of Becker \&
Butler (\cite{Bec88}, \cite{Bec89}, \cite{Bec90}). The atomic data used for 
C\,{\sc ii} are from Eber \& Buter (\cite{Ebe88}), and the Mg\,{\sc ii}
data are from Mihalas (\cite{Mih72}).

\begin{table}[b]
\caption[]{Non-LTE atmospheric parameters for OB78-277 and its
Galactic standard, HD\,13854. Wind parameters of OB78-277 are adopted to
be identical with those of HD\,13854}
\begin{flushleft}
\centering
\begin{tabular}{llcc} 
\hline \hline
Parameter & &OB78-277   & HD\,13854   \\
\hline      
\\            
$T_{\rm eff}$ &(K)                    & 21,000        & 21,000      
\\ 
$\log$ g      & (dex)                 &  2.4          &  2.4	    
\\     
$\xi$         &(kms$^{-1}$)           &  16           &  17
\\
v$_{\infty}$  &(kms$^{-1}$)           & 1000          & 1000
\\
\.{M}         &(M$_{\odot}$ yr$^{-1}$)& 0.3*10$^{-6}$ & 0.3*10$^{-6}$
\\  
$\beta$-parameter&                    & 3.00          & 3.00
\\ 
R$_{*}$       &(R$_{\odot}$)          & 40.8          & 40.8
\\      
\hline 
\end{tabular}
\end{flushleft}
\label{nltepar} 
\end{table}

\begin{figure}
\resizebox{\hsize}{!}{\includegraphics{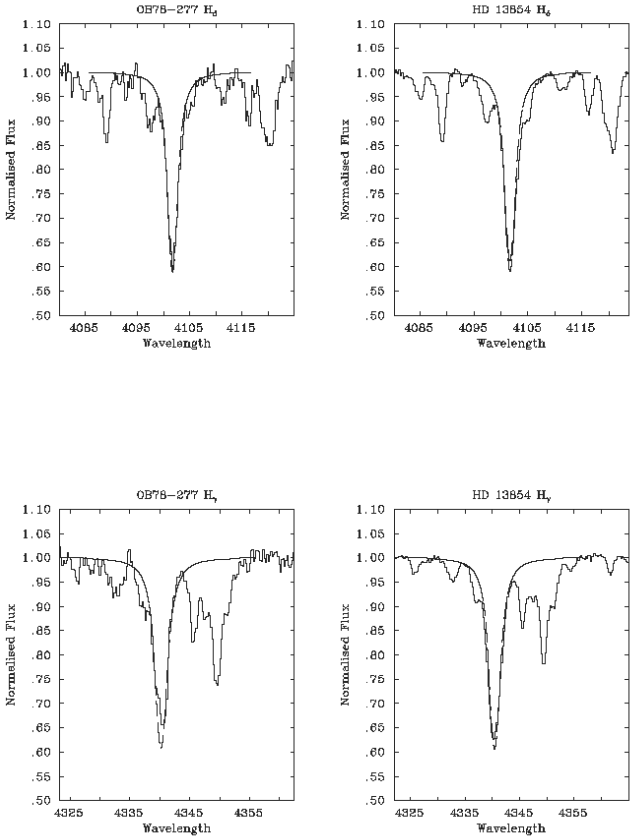}}
\caption[]{\label{nlteloggs}The H$_{\delta}$ and H$_{\gamma}$ profiles for OB78-277 and
its Galactic counterpart, HD\,13854. The theoretical profiles
have surface gravities of $\log$ g = 2.4 and are represented
by the dashed (- -) lines.The observed spectra are represented
by the solid lines.
}
\end{figure}

The effective temperature and surface gravity of OB78-277 and
its standard, HD\,13854, were estimated using similar procedures
to that discussed in Sect.~\ref{lteatpar} i.e from the Si {\sc
iii}/ Si {\sc iv} ionisation balance and profile fitting of
H$_{\delta}$ \& H$_{\gamma}$. In Fig.~\ref{nlteloggs}, we have
plotted theoretical profiles of H$_{\delta}$ \& H$_{\gamma}$ for
$\log$ g = 2.4 over normalised observed spectra of OB78-277 and
HD\,13854. From a comparison with Fig.~\ref{lteloggp}, it is
clear that the wings of the non-LTE profiles fit the data better
than in the LTE profiles for both of the Balmer lines. The
cores of the non-LTE theoretical profiles are not as deep as the
LTE profiles and also fit the observed data more accurately.
This is due to the treatment of the stellar wind in the non-LTE
analysis.

\begin{table*}
\caption[]{Non-LTE absolute and differential elemental abundances
of OB78-277 and its Galactic standard, HD\, 13854. Also included
are the results of the LTE differential analysis for
comparison. These have been recalculated using  the same
lines as used in the NLTE analysis only. $\Delta[\frac{X}{H}] = \log
[\frac{X}{H}]_{*} - \log[\frac{X}{H}]_{standard}$}
\begin{flushleft}
\centering
\begin{tabular}{lllll} 
\hline \hline
ION &\multicolumn{2}{c}{12 + log [$\frac{X}{H}$]} &\multicolumn{2}{c}{$\Delta[\frac{X}{H}]$}\\
    & OB78-277   & HD\,13854   & non-LTE   &  LTE \\
\hline      
\\            
C {\sc ii}   & 7.76             & 7.56            & +0.20             & +0.21   
\\
N {\sc ii}   & 7.85 $\pm$ 0.14  & 7.95 $\pm$ 0.12 & +0.03 $\pm$ 0.09  & -0.11 $\pm$ 0.05    
\\     
O {\sc ii}   & 8.71 $\pm$ 0.08  & 8.33 $\pm$ 0.22 & +0.38 $\pm$ 0.11  & +0.23 $\pm$ 0.10
\\
Mg {\sc ii}  & 7.55             & 7.65            & -0.10             & +0.08  
\\     
Si {\sc iii} & 7.42 $\pm$ 0.17  & 7.14 $\pm$ 0.05 & +0.28 $\pm$ 0.02  & +0.19 $\pm$ 0.05 
\\  
Si {\sc iV}  & 7.34             & 7.09            & +0.25             & +0.08   
\\     
\hline
\end{tabular}
\end{flushleft}
\label{nlteabund} 
\end{table*}

The wind terminal velocity (v$_{\infty}$) for HD\, 13854 was taken from
Kudritzki et al. (\cite{Kud99}), whereas the mass-loss rate (\.{M}) and
$\beta$-parameter were determined from the H$_{\alpha}$ profile. The
stellar radius was derived from the absolute magnitude, deduced from
the apparent magnitude after adopting a distance modulus (Garmany \&
Stencel \cite{Gar92}), and the emergent flux of the model atmosphere
(Kudritzki 1987). As the wind parameters of HD\,13854 describe an
intermediate density wind there is no contamination of the weak
metallic lines which are formed deep enough in the stellar atmosphere
to be hardly effected by the wind. We have therefore adopted the same
values of \.{M}, v$_{\infty}$ and $\beta$-parameter for OB78-277 as
that of its Galactic counterpart. The microturbulence and silicon
abundance were determined simultaneously from the Si {\sc iii}
multiplet at $\sim$ 4560 \AA . The atmospheric parameters of both
OB78-277 and HD\, 13854 are shown in Table~\ref{nltepar}. Adopting the
microturbulence estimated from the Si {\sc iii} lines the abundances
of carbon, nitrogen, oxygen and magnesium were obtained (see
Table~\ref{nlteabund}). 

In Table~\ref{nlteabund} we present the non-LTE absolute and
differential abundance results and include the LTE differential
abundances, recalculated for the same lines as used in the
non-LTE analysis. We can see the LTE and non-LTE {\em
differential} abundance estimates differ by a maximum of $\pm$
0.18 dex. This is reassuring as it is smaller than the
uncertainties in our LTE abundance analysis and hence we can
believe that the LTE differential analysis is not predominantly
affected by non-LTE effects. Furthermore the difference in {\em
absolute} oxygen abundances between the LTE and non-LTE analyses
is 0.07 dex (see Table~\ref{m31diff} \& \ref{nlteabund}), and
therefore smaller than our estimated errors in the LTE analysis.
In Fig~\ref{abund_Rg} we included the non-LTE oxygen abundance
of OB78-277, it is clear that if non-LTE effects of the same
magnitude were present for the other B-type supergiants there
would be no significant effect on the derived abundance gradient
of M31.

In Sect.~\ref{ltemod} we have discussed the LTE analysis of OB10-64 (see
Table~\ref{m31ab}), this star has been analysed previously by
Smartt et al. (\cite{Sma01b}) using a non-LTE model atmosphere code. The
model atmospheres generated by Smartt et al. (\cite{Sma01b}) used the
plane-parallel, non-line blanketed TLUSTY code of Hubeny, 1988.
For the line formation calculations the codes DETAIL (Giddings
\cite{Gid81}) and SURFACE (Butler \cite{But84}) were used. It is reassuring to
note that comparing the non-LTE differential analysis of OB10-64
from Smartt et al. (\cite{Sma01b}) to the LTE analysis included here, a
maximum difference in the abundances of $\pm$ 0.18 dex is again
observed. Moreover comparing the non-LTE absolute oxygen
abundance of Smartt et al. (\cite{Sma01b}) to the LTE abundance given in
Sect.~\ref{ltemod} a difference of 0.05 dex is found (see Fig~\ref{abund_Rg}).

In conclusion, the results of this section imply that the non-LTE
effects on the abundances in these B-type supergiants appear to be
less significant than the typical errors in our LTE analysis. 
Morover, the results from the differential LTE
analyses appear to be reliable indicators of the chemical
composition in these M31 supergiants.

%-----------------------------------------------------------------------
%
\section{Discussion} 
\label{discuss}
Here we first discuss the degree of processing in each of the
M31 supergiants followed by a comparison of the abundances
determined in this analysis with those of previous nebular and
stellar investigations. Finally the abundance gradient of M31
will be discussed in detail. 

\subsection{Degree of chemical processing in the M31 supergiants} 
\label{chempro}

Chemical peculiarities can be present in stellar
photospheres due to mixing of core-processed material.
Current models predict that if CNO-cycled material was mixed
to the surface there would be an observed enhancement in
helium and nitrogen accompanied by a depletion in carbon and
oxygen (see Maeder \& Meynet \cite{Mae88}; Gies \& Lambert
\cite{Gie92}; Meynet \& Maeder \cite{Mey00}; Heger \& Langer
\cite{Heg00}).  In a sample of 46 Galactic B-type
supergiants MLD identified three sub-groups based on the
amount of processing observed in the stellar photosphere:
`normal/moderate', `processed?' and `highly processed'. The
first group could be associated with stars having chemically
near normal photospheres whilst the latter is designated to
stars which clearly show products of CN-cycled material in
their photospheres. The `processed?' group may have some
contamination from core material but of a lesser extent than
the 'highly-processed' stars. In their analysis MLD
concluded that there was no obvious correlation between
helium abundances and the assigned chemical sub-groups.
However in the most luminous stars and those in the `highly
processed' category oxygen had stronger features than the
chemically normal supergiants. Thus no conclusive
information on the correlation of oxygen with the degree of
processing could be obtained.

By comparing the carbon and nitrogen line strengths and
abundances of the M31 supergiants to those of their Galactic
counterparts and considering the labels assigned by MLD, we
shall discuss to what degree the M31 stars are processed, if at
all. Smartt et al. (\cite{Sma01b}) suggested that OB10-64 belongs to the
'processed?' sub-group, this is consistent with our findings.
This star has similar carbon and nitrogen line strengths as that
of the Galactic supergiant HD\,~167264, which has been classified
as 'processed?' by MLD. 

\begin{figure}
\resizebox{\hsize}{!}{\includegraphics{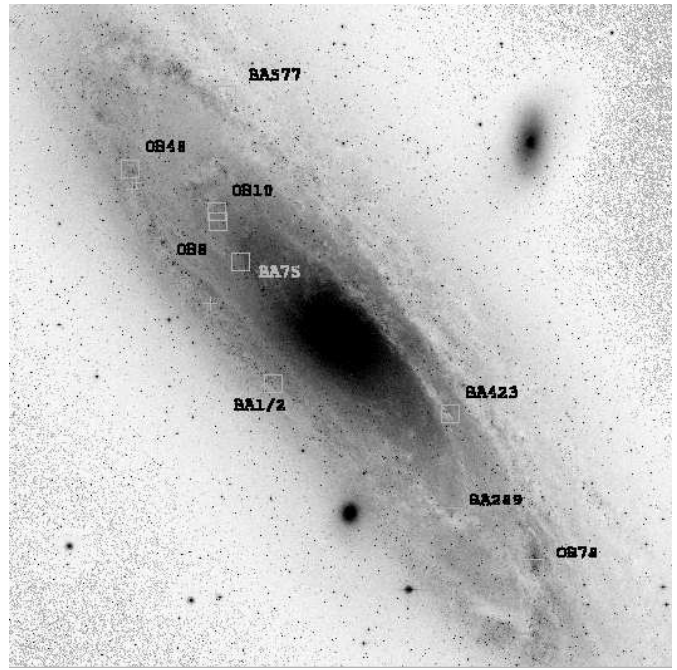}}
\caption[]{\label{XYpositions} Mosaic of DSS images of M31 with the
positions of associations OB8, OB10, OB48 and OB78 represented by
squares. The image size is 90'$\times$90' with north up
and east to the left. The three pluses (+) are the  A-type supergiants
from VMLPKL in increasing order of DEC; 41-2368, 41-3712 \&
41-3654. The F-supergiant (A-207) from the VMLPKL sample lies off
the field-of-view in the outer south west corner of M31. The H {\sc
ii} regions from Blair et al. (1982) which fall in this
90'$\times$90' image are displayed. The H\,{\sc ii} regions K315 \&
K703, of Galarza et al. (1999), have been omitted for clarity of
the image. K315 is coincident with the OB10 association, whereas
K703 lies very close but not coincident with the OB48
association.}  
\end{figure}

As the standard for OB48-358 is classified as
`normal' the differential analysis is easier to interpret.
OB48-358 appears to be in the `normal/moderate' category as the
C{\sc ii} 4267 \& N {\sc ii} 3995 lines strengths are similar to
that of the standard. Although the C{\sc ii} 4267 line of OB8-76
is of similar line strength to that of its standard the N{\sc ii}
is stronger indicating that `processed?' or `highly-processed'
would be the appropriate classification. MLD plotted the N/C line
strength ratios against the logarithmic effective temperature of
their Galactic supergiants and found that there is a correlation
between the degree of chemical processing and this parameter (see
Fig 9. ~MLD). From this plot it would appear that OB8-76 is a
member of the `processed?' sub-group. The two OB78 supergiants
appear to be either `processed?' or `highly processed'. In both
stars the carbon line strengths are similar to their `highly
processed' standards. For OB78-277 the nitrogen is slightly weaker
than that of its Galactic counterpart and from its line strength
ratio it appears to be `processed?', whilst all the evidence
points to OB78-478 being a member of the `highly processed' group.
OB8-17 appears to be as processed as HD\,13584 if not slightly more
as indicated from the N {\sc ii} 4630 \AA\ line. As HD\,13854 is in
the `highly processed' sub-group OB8-17 must also belong to this
classification. In OB48-234 carbon is stronger than the standard
while nitrogen is weaker purely indicating that this star belongs
to the `normal/moderate' group as the standard is `highly
processed. These classifications are summarised in
Table~\ref{m31par}.

%-----------------------------------------------------------------------
%
\subsection{Comparison of B-type supergiants with previous results.}
\label{prev}

\begin{figure*} 
\vbox{\epsfxsize=12cm\epsfbox{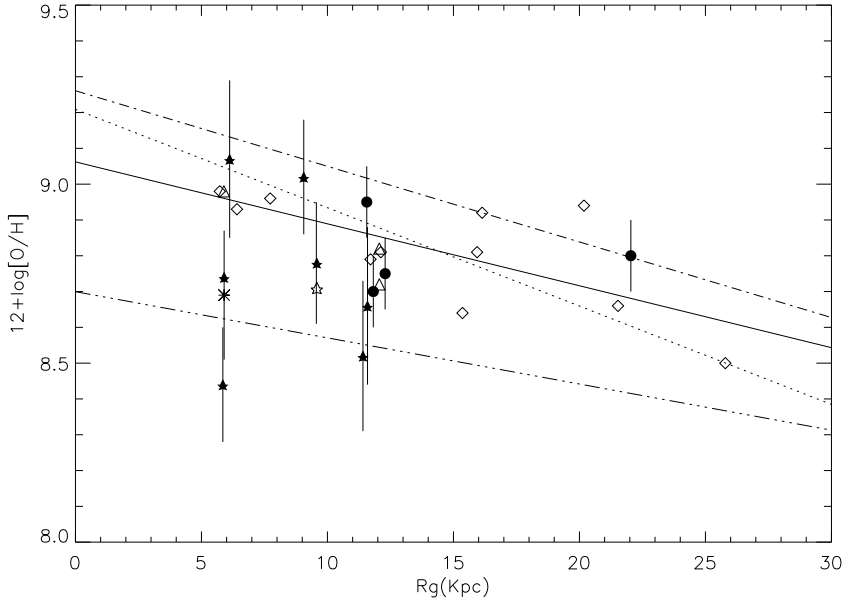}\vspace{-8.5cm}}
\hfill\parbox[b]{5.5cm}{\caption[]{\label{abund_Rg}Oxygen abundances
for H\,{\sc ii} regions and supergiants in M31 as a function of
galactocentric distances. The solid stars ($\star$) are the abundance
results of the seven B-type  supergiants in this study. The unfilled
star is the non-LTE oxygen abundance for OB78-277. The asterisk
($\ast$) is the photospheric NLTE abundance of OB10-64 from Smartt et
al. (\cite{Sma01b}). The open diamonds ($\diamond$) and open triangles
($\triangle$) are the abundances of the H\,{\sc ii} regions from Blair
et al. (\cite{Bla82}) and Galarza et al. (\cite{Gal99}), respectively,
calculated using McGaugh (\cite{McG91}) calibrations. The solid
circles ($\bullet$) are the abundances of four A-F-type supergiants
from VMLPKL. The dotted ($\cdot\cdot\cdot$), solid (-), dash-dot
(-$\cdot$-) and dash-dot-dot-dot(-$\cdot\cdot\cdot$-) lines are the
least squares fit  through the H{\sc ii} regions of Blair using Mc
Gaugh (\cite{McG91}), Pagel et al. (\cite{Pag80}), Zaritsky et al.
(\cite{Zar94}) and Pilyugin (\cite{Pil01a}) calibrations,
respectively. The error bars in the stellar results represent the
standard error in the mean.}} 
\end{figure*}

To compare the abundances of this set of supergiants with those
of previous stellar studies, it is appropriate to discuss the
composition of stars at similar galactocentric distances. Only one
previous stellar investigation by VMLPKL has been carried out in
M31 (the results of Smartt et al. \cite{Sma01b} have been subsumed into
this paper). Two of their M31 A-type supergiants lie in close
proximity to the OB48 association at a galactocentric distance
of approximately 11.5 kpc (see Fig.~\ref{XYpositions} \&
\ref{abund_Rg}). The common elements observed in the A \&
B-type supergiants are Fe and the $\alpha$-processed elements
Si, O, Mg. It is important to note that the A-type supergiants
were analysed using a non-LTE analysis, hence when comparing
abundances between the A \& B-type supergiants the uncertainties
discussed in Sect.~\ref{nlte} must be considered. The $\alpha$-processed
element abundances in these A \& B-type supergiants are in good
agreement, {\it within the errors}. The OB48 oxygen abundances
are $\sim$ 0.1 dex lower than those of the A-type supergiants but
given the errors of both analyses, including the non-LTE effects,
this difference is considered to be negligible. OB48-234 has a
marginal \ion{Fe}{iii} feature which gives a significantly lower
Fe abundance than that observed in the A-type supergiants. However
given that this feature is weak and that the iron abundances
from OB8-17 \& OB78-478 are similar to that of the A-type
supergiants, no significance is given to this result.

From nebular studies only abundances of oxygen are available
for comparison with our results. In addition to the two
A-type supergiants near association OB48, there are also two H {\sc
ii} regions from the analysis of Galarza et al. (\cite{Gal99}) (K703 \&
K722). Unfortunately these H\,{\sc ii} regions are not spatially coincident
with our targets. There appears to be a 0.2 dex difference in the
oxygen abundances derived from the H\,{\sc ii} regions and those
from the B-type supergiants, with the former having the higher
abundances. However, this depends on the calibration of the
line ratio parameter, R$_{23}$ with [O/H], and the various
calibrations give a spread of $\sim$ 0.2 dex in abundance.
For example using the calibration of Mc Gaugh (\cite{McG91}), the
magnitude of the difference in abundance between the two H {\sc
ii} regions mentioned above and the OB48 stars is 0.2 dex, whereas
adopting the Zaritsky et al. (\cite{Zar94}) calibration a difference
of 0.36 dex is observed (see Fig.\ref{abund_Rg}). Smartt et al.
(\cite{Sma01b}) also observed this difference in the abundances derived
from these calibrations of the R$_{23}$ parameter when comparing
the abundances of OB10-64 to that of a spatially coincident H
{\sc ii} region, K315 (Galarza et al. \cite{Gal99}). In contrast to this, VMLPKL found that
there was good agreement of the oxygen abundances in H {\sc
ii} regions with that of their A-type supergiant sample.

The above mentioned calibrations of the R$_{23}$ with [O/H] all
have a systematic error inherent in their results, as they have not
considered the physical conditions (i.e. the hardness of the
ionizing radiation and the geometrical conditions) of the H\,{\sc ii}
regions (for further details see Pilyugin \cite{Pil00}). A
more recent calibration by Piyugin (\cite{Pil01a}, \cite{Pil01b}), known as the
P-method, considers the physical conditions of the H\,{\sc ii} regions
by introducing an excitation index, P. This excitation index has a
correlation with the effective temperature of the ionizing star and
hence is a good indicator of the hardness of the ionizing
radiation. The oxygen abundances derived for the M31 H\,{\sc ii}
regions using this P-method are significantly less ($\sim$ 0.3-0.5 dex) than
those derived using the calibrations of Mc Gaugh (\cite{McG91}) \& Zaritsky
et al. (\cite{Zar94}) and are in better agreement with the oxygen abundances
of the B-type supergiants in our sample. Furthermore  using this
calibration a solar oxygen abundance is indicated for the center of
M31 (see Fig ~\ref{abund_Rg}).

%-----------------------------------------------------------------------
%
\subsection{Abundance gradients and ratios of M31}
\label{grad}

The form of the abundance gradient in M31 is
still open to debate. Nebular studies in recent years have
suggested that a shallow abundance gradient is present in
M31, whereas the one stellar investigation (of just three
stars) does not show a clear gradient (Blair et al.
\cite{Bla82}; Galarza et al. \cite{Gal99}; VMLPKL). 

Taking the set of 11 H\,{\sc ii} regions of Blair et al.
(\cite{Bla82}), we have plotted the oxygen abundance as a
function of the M31 galactocentric distance (see
Fig.~\ref{abund_Rg}). The plotted data points (open diamonds,
$\diamond$) are those derived using the calibration of Mc Gaugh
(\cite{McG91}), with the galactocentric distances recalculated for the 
latest value of the distance to M31; 783 $\pm$ 30 kpc from
Holland (\cite{Hol98}). The least-squares fit to these data
points gives an oxygen abundance gradient of -0.027 dex
kpc$^{-1}$ and implies from extrapolation a central oxygen
abundance of ~9.21 dex. Also plotted in Fig.~\ref{abund_Rg} are
the least-squares fit to the nebular oxygen abundances adopting
the calibrations of Zaritsky et al. (\cite{Zar94}), Pagel et al.
(\cite{Pag80}) and Piyugin (\cite{Pil01a}) . These imply
shallower gradients of -0.021 dex kpc$^{-1}$, -0.017 dex
kpc$^{-1}$ and -0.013 dex kpc$^{-1}$ and central oxygen
abundances of 9.20 dex, 9.06 dex and 8.70 dex, respectively.  
These results are summarised in Table\,\ref{gradients} for easy
comparison with the B-supergiant results.

\begin{table}
\caption[]{Comparison of the oxygen abundance gradients derived from 
the H\,{\sc ii} region data of Blair et al. (1982) using 
four different calibrations, and the stellar results from 
A and B-type supergiants.}
\begin{flushleft}
\centering
\begin{tabular}{llll} 
\hline \hline
Elment & Data   & Calibration & Gradient (dex kpc$^{-1}$)\\ \hline
Oxygen & H{\sc ii} regions & McGaugh         & $-0.027 \pm 0.01$\\
Oxygen & H{\sc ii} regions & Zaritsky et al. & $-0.021\pm 0.01$\\
Oxygen & H{\sc ii} regions & Pagel et al.    & $-0.017 \pm 0.01$\\
Oxygen & H{\sc ii} regions & Pilyugin        & $-0.013 \pm 0.01$\\
\\
Oxygen    &  B-type stars     & This paper      & $-0.006 \pm 0.02$ \\
Magnesium & B-type stars     & This paper      & $-0.023 \pm 0.02$ \\
Silicon   &  B-type stars     & This paper      & $-0.009 \pm 0.02$ \\
\hline
\end{tabular}
\end{flushleft}
\label{gradients} 
\end{table}

The least-squares fit to our sample of supergiants within 12 kpc
of the galactic center suggests an abundance gradient consistent
with that derived from the nebular results, -0.017 $\pm$ 0.02
dex kpc$^{-1}$. However as OB8-76 is possibly a multiple
system, this result should be omitted from our abundance
gradient analysis. This then gives a negligible oxygen 
abundance gradient in M31 of -0.06 $\pm$ 0.02 dex kpc$^{-1}$. Furthermore negligible abundance
gradients are also derived from the average silicon abundances
in each of the four OB associations. However from the magnesium
abundances in each association an abundance gradient of -0.023
$\pm$ 0.02 dex kpc$^{-1}$ is observed. VMLPKL also investigated
the radial oxygen abundance gradient of M31, unlike the nebular
results, they found that there is no gradient between 10 kpc and
20 kpc. However as the VMLPKL study has abundances
from only three objects over a fairly narrow galactocentric
range, one must take this as a very preliminary result. 

As discussed in Sect. ~\ref{chempro} variations in the
photospheric abundances of carbon, nitrogen and oxygen from
their initial values have been predicted by stellar
evolutionary models of blue supergiants, such as those based on
stellar rotation (see Meynet \& Maeder \cite{Mey00}; Heger \&
Langer \cite{Heg00}). However these models predict that only
very small changes in the oxygen abundance will have taken
place by the end of the main-sequence ($\sim$ 10\% by number).
Moreover such a depletion in oxygen has not been
observationally proven in B-type supergiants which have highly
processed photospheres (for example see MLD). If there is a
change in oxygen of the magnitude predicted by these models it
is masked by the uncertainties in our observations and
therefore would not effect the oxygen abundance gradient.
Reassuringly, the silicon and magnesium gradients are consistent
with that of the oxygen abundance gradient and these elements
should not be effected by the products of CNO-cycle burning.

The O, Mg, Si \& S abundances of OB8-17, the innermost star of our
sample, are comparable to those of its Galactic counterpart;
HD\,13854. This is also the case for the nearby star
OB10-64, as its O, Mg and Si abundance are very similar to that of
the standard star HD\,167264. This implies that unlike the common
opinion that the center of M31 is metal rich, as indicated by
nebular results (Blair et al. \cite{Bla82}; Galarza et al. \cite{Gal99}), it
appears to be of solar metallicity. This was suggested by Smartt et
al. (\cite{Sma01b}) from their NLTE analysis of OB10-64 and is consistent
with the nebular results calculated using the P-method calibration
of Pilyugin (\cite{Pil01a}, \cite{Pil01b}). 

At this point it is worthwhile discussing the element ratios
[O/Fe], [Mg/Fe] \& [Si/Fe] in M31. The ratio of these
$\alpha$-processed elements to iron can indicate the star
formation history of M31 due to the different
nucleosynthetic origins of these elements. The bulk of  iron
production is thought to come from low and intermediate mass
stars in degenerate binary systems during explosions of 
Type\,Ia supernovae, whereas the $\alpha$-processed elements
are mainly produced in massive stars which enrich the ISM
through core-collapse Type\,II supernovae. VMLPKL observed
an increase in $\Delta$[O/Fe] toward the outer disk of M31,
and noted a possible increase in $\Delta$[$\alpha$/Fe],
excluding oxygen. In Fig~\ref{X_Fe} we plot the element
ratios $\Delta$[O/Fe], $\Delta$[Mg/Fe] \& $\Delta$[Si/Fe]
against galactocentric distance for our target sample and
for that of VMLPKL. Consistent with the findings of VMLPKL,
there is evidence for an increasing gradient in each of
these elements.

{\begin{figure}
\resizebox{\hsize}{!}{\includegraphics{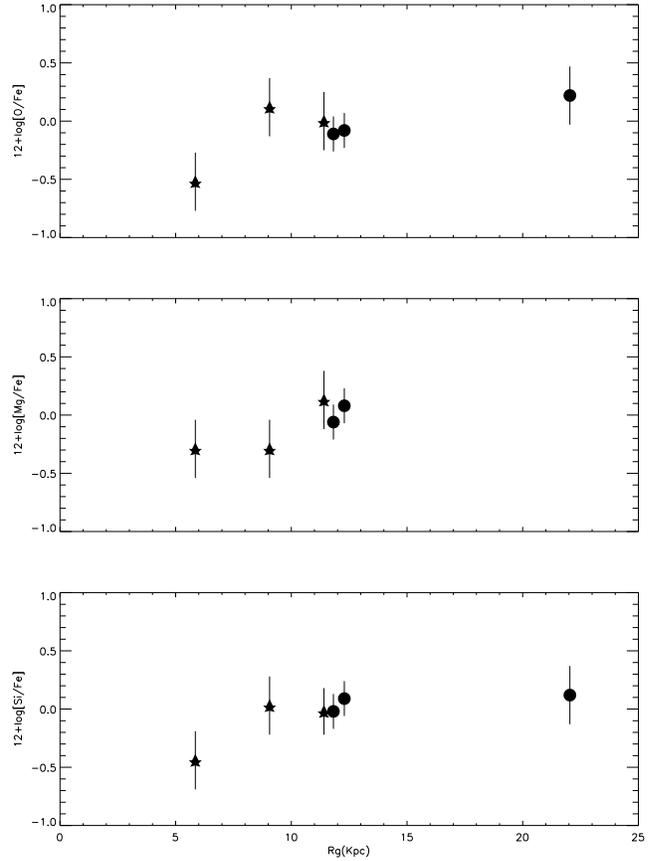}}
\caption[]{\label{X_Fe} The element ratios $\Delta$[O/Fe],
$\Delta$[Mg/Fe] and $\Delta$[Si/Fe] from our B-type supergiant
sample ($\star$) and the A-F-type supergiants of VMLPKL ($\bullet$).
Where $\Delta$[X/Fe] = [X/Fe]$_{\rm M31}$ - [X/Fe]$_{\rm
Standard}$, in the case of the B-type supergiants the standards are as
those for the differential analysis (see Table~\ref{m31diff}),
whereas for the A-F-type supergiants the sun was taken as a standard.}
\end{figure}

%-----------------------------------------------------------------------
%
\section{Conclusion}
\label{conc}

We have presented the results of detailed LTE absolute and
differential abundance analyses of the largest group of B-type
supergiants in M31 studied to date.  Non-LTE calculations have
also been carried out to investigate the effect of departures
from LTE on our results. It was shown that although
analysed using LTE model atmospheres and line formation codes
the analysis is dominated by uncertainties stemming from the
quality of the data rather than non-LTE effects.

The seven B-type supergiants lie in distinct clusters which
cover a galactocentric distance of 5$-$12 kpc and from the
derived abundances we estimated the oxygen abundance gradient of
M31. Across this fairly restricted range of the disk we do not
find any evidence of a significant abundance gradient, which is
similar to the result found by VMLPKL for four A-F -type
supergiants.  Radial abundance gradients for the
$\alpha$-processed elements, Si \& Mg, were also determined
indicating a negligible silicon abundance gradient and a
possible shallow gradient for magnesium.  However we emphasise
that we still have a very restricted number of data points with
which to probe the abundance gradient in M31. The work of Smartt
\& Rolleston (\cite{smr97}), in determining abundance gradients
in the Milky Way, cautions against using small numbers of stars
with restricted Galactocentric radii to draw firm conclusions on
existance of abundance gradients. This result is reasonably
consistent with the shallow  negative oxygen abundance gradients
determined from H\,{\sc ii} regions and supernovae remnants
(Dennefeld \& Kunth \cite{Den81}; Blair et al. \cite{Bla82};
Galarza et al. \cite{Gal99}). 

It has been shown that dependent on which empirical calibration
of the R$_{23}$ parameter with [O/H] that is adopted, different
magnitudes of the radial abundance gradient in M31 are obtained.
It has also been shown that there is an offset between the
abundances obtained from the H\,{\sc ii} regions and that of the
B-type supergiants ($\sim$ 0.15-0.4 dex), at the same
galactocentric distance, which again depends on the empirical
calibration implemented. These results indicate that the
calibration of the R$_{23}$ parameter with [O/H] for H\,{\sc ii}
regions with high metallicity/low excitation are clearly not
accurately constrained. The empirical calibration which fits our
stellar results best is that of Pilyugin (\cite{Pil00},
\cite{Pil01a}, \cite{Pil01b}). Moreover this calibration gives
the shallowest oxygen abundance gradient ($\sim$ 0.013 dex
kpc$^{-1}$). The main difference between this and other
calibrations is that it considers the hardness of the ionising
radiation and therefore the physical parameters of the H\,{\sc
ii} regions. As the B-supergiants only probe out to 12\,kpc
from  the centre of M31, it would be highly desirable to  sample
the outer regions of the M31 disk with similar stars. 

Our results for the innermost supergiants of the sample, OB8-17
\& OB10-64, indicate that M31 is not metal-rich, as previously
thought from the results of H\,{\sc ii} regions, but actually
suggests that it is of solar metallicity. This is consistent with
the results obtained by Smartt et al. (\cite{Sma01b}) 
and from the nebular
oxygen abundances when implementing the empirical calibration of
Pilyugin (2000, 2001a, 2001b). Although the star OB8-76 
appears to have quite high abundances, the mean abundance of the 
OB8 cluster is not significantly above the solar neighbourhood. 

Smartt et al. (\cite{Sma01b}) have shown that detailed wind-analyses can be
accurately carried out on B-type supergiants in M31. They also
found that the wind momentum-luminosity relation of Kudritzki et
al. (\cite{Kud99}) can be applied to these B-type supergiants. In the
future to calibrate the WLR for the metallicity of M31 and further
the calibration of WLR in the Local Group, the atmospheric
parameters and abundances will be used in conjunction with the
terminal velocity (see Bresolin et al. \cite{bres2002})
and mass-loss rates of the wind. The final aim
of this work will be to provide an accurate and independent
extragalactic distance scale.

%-----------------------------------------------------------------------
%
\section*{Acknowledgements}   

We are grateful to the continuous support from A.  Herrero of
the IAC, Spain. We thank J. Puls for providing the non-LTE
'unified model atmosphere' code. CT is grateful to the
Department of Higher and Further Education, Training and
Employment for Northern Ireland (DEFHTE) and the Dunville
Scholarships fund for their financial support. PLD is grateful
to the UK Particle Physics \& Astronomy Research Council
(PPARC) for financial support. SJS also thanks PPARC for
financial support in the form of an Advanced Fellowship award. 
The WHT is operated on the island of La Palma by the Isaac
Newton Group in the Spanish Observatorio del Roque de los
Muchachos of the Instituto de Astrof\'{i}sica de Canarias.

%-----------------------------------------------------------------------
%

\end{document}